\documentclass[conference]{IEEEtran}
\IEEEoverridecommandlockouts

\usepackage{cite}
\usepackage{amsmath,amssymb,amsfonts,amsthm}
\usepackage{algorithmic}
\usepackage{graphicx}
\usepackage{textcomp}
\usepackage{xcolor}
\def\BibTeX{{\rm B\kern-.05em{\sc i\kern-.025em b}\kern-.08em
    T\kern-.1667em\lower.7ex\hbox{E}\kern-.125emX}}

\usepackage{mathtools}
\usepackage{physics}
\usepackage{dsfont}
\renewcommand\vec{\mathbf}
\newcommand{\ketf}[1]{|{#1})}

\newcommand{\m}[1]{\mathfrak{#1}}
\usepackage{comment}
\usepackage{algorithm}

\usepackage{makecell}
\usepackage{booktabs}

\theoremstyle{definition}

\newtheorem{theorem}{Theorem}

\newcommand{\iu}{\mathrm{i}\mkern1mu}
\newcommand{\B}[1]{\textbf{#1}}

\usepackage{multirow}
\usepackage{subcaption}

\usepackage{hyperref}
\usepackage{cleveref}

\begin{document}

\title{Optimal fermion-qubit mappings via quadratic assignment
\thanks{This work was in part supported by NSF award \#2427042 and DARPA ONISQ program. M.L.C. received the support of a Cambridge Australia Allen \& DAMTP Scholarship during the preparation of this work. S.S. was supported by the Royal Society University Research Fellowship and “Quantum simulation algorithms for
quantum chromodynamics” grant (ST/W006251/1).}
}

\author{
\IEEEauthorblockN{Mitchell Chiew\IEEEauthorrefmark{1},
Cameron Ibrahim\IEEEauthorrefmark{2},
Ilya Safro\IEEEauthorrefmark{2}, and
Sergii Strelchuk\IEEEauthorrefmark{3}}
\IEEEauthorblockA{\IEEEauthorrefmark{1}\textit{Department of Applied Mathematics and Theoretical Physics} \\
\textit{University of Cambridge
Cambridge, United Kingdom}\\
mlc79@cam.ac.uk}
\IEEEauthorblockA{\IEEEauthorrefmark{2}\textit{Department of Computer and Information Sciences} \\
\textit{University of Delaware
Newark DE, USA} \\
\{cibrahim, isafro\}@udel.edu}
\IEEEauthorblockA{\IEEEauthorrefmark{3}\textit{Department of Computer Science} \\
\textit{University of Oxford
Oxford, UK }\\
sergii.strelchuk@cs.ox.ac.uk}
}

\maketitle
\begin{abstract}
Simulation of fermionic systems is one of the most promising applications of quantum computers. It spans problems in quantum chemistry, high-energy physics and condensed matter. Underpinning the core steps of any quantum simulation algorithm, \textit{fermion-qubit mappings} translate the fermionic interactions to the operators and states of quantum computers. This translation is highly non-trivial: a burgeoning supply of fermion-qubit mappings has arisen over the past twenty years to address the limited resources of early quantum technology. Previous literature has presented a dichotomy between ancilla-free fermion-qubit mappings, which minimise qubit count, and local encodings, which minimise gate complexity.

We present two computational approaches to the construction of general mappings while working with a limited number of qubits, striking a balance between the low-qubit and low-gate demands of present quantum technology. The first method frames the order of fermionic labels as an instance of the quadratic assignment problem to minimize the total and maximum Pauli weights in a problem Hamiltonian. 
We compare the order-optimized performance of several common ancilla-free mappings on systems of size up to 225 fermionic modes.
The second method is a computational approach to incrementally add ancilla qubits to Jordan--Wigner transformations and further reduce the Pauli weights.
By adding up to 10 ancilla qubits, we were able to reduce the total Pauli weight by as much as 67\% in Jordan--Wigner transformations of fermionic systems with up to 64 modes, outperforming the previous state-of-the-art ancilla-free mappings.

\noindent Reproducibility: source code and data are available at  \url{https://github.com/cameton/QCE_QubitAssignment}
\end{abstract}

\begin{IEEEkeywords}
fermion-qubit mapping, quantum computing, ancilla qubits, quadratic assignment.
\end{IEEEkeywords}

\section{Introduction}

\begin{figure*}
    \centering
    \includegraphics[width=0.9\linewidth]{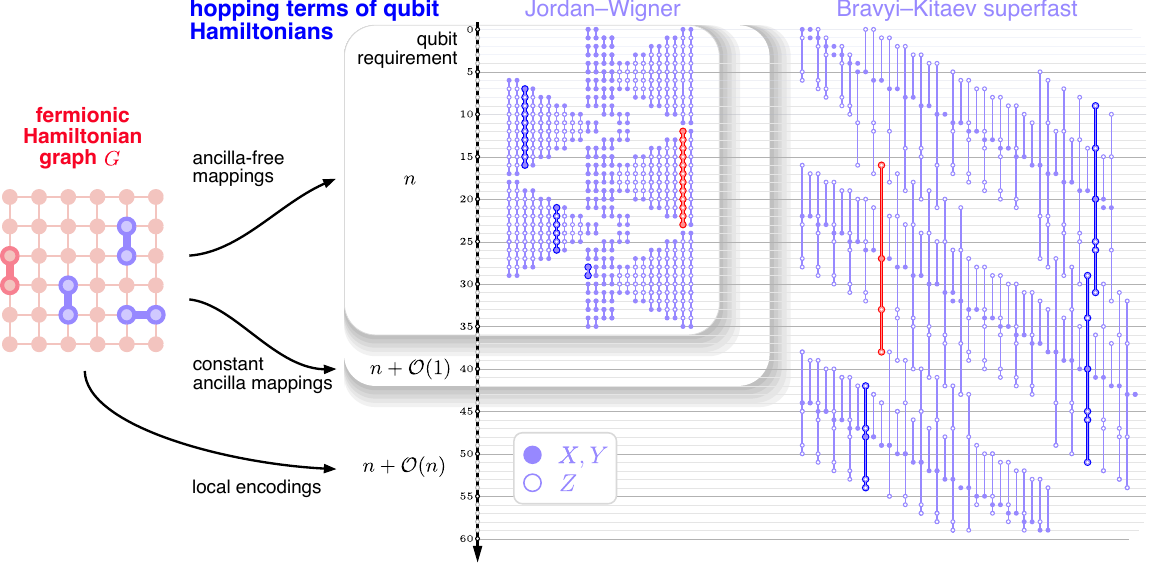}
    \caption{Demonstration of the differing localities of fermion-qubit mappings. Ancilla-free mappings such as the Jordan--Wigner transformation produce hopping terms that act nonlocally, i.e.\ on a number of qubits that increases with system size, while local encodings such as the Bravyi--Kitaev superfast transformation employ ancilla qubits to produce local hopping terms which act on a constant number of qubits. Each vertical line on the qubit register depicts a hopping term in the qubit Hamiltonian, acting with Pauli $X$, $Y$ or $Z$ matrices on different qubits as indicated by the circles.}
    \label{fig:different_mappings}
\end{figure*}

The study of second-quantized fermionic systems in condensed matter physics and quantum chemistry presents a significant challenge to the computational sciences. Hamiltonian simulation and spectral analysis such as ground-state energy estimation of fermionic systems remains an exponentially scaling problem for classical computers \cite{kempe_complexity_2006, mcclean2014exploiting, ogorman_electronic_2021}. Quantum computing provides a potential advantage through alternative strategies such as Trotterization \cite{lloyd1996universal}, quantum signal processing \cite{low2017optimal} and variational quantum eigensolvers \cite{peruzzo2014variational}, limited by the fidelity and scale of contemporary quantum technology.

Due to the lack of straightforward efficient  representations of fermionic interactions in the standard circuit model, any attempt must begin with a \textit{fermion-qubit mapping}.
The complexity of modelling evolution and extraction of key information about a fermionic system hinges on the complexity of such representations. A central factor is the \textit{locality} of the qubit Hamiltonian, relating the qubit number or geometry of the support of each term to the task of simulation \cite{berry2015hamiltonian} or ground-state estimation \cite{peruzzo2014variational}.
The most well-known mapping, the Jordan--Wigner transformation \cite{jordan:1928wi}, is conceptually intuitive but leads to representations of local fermionic excitations with supports that scale linearly in system size. For example, all qubits on the quantum register may need to be engaged simultaneously to represent an electron hopping between adjacent molecular orbitals.
The non-locality of the Jordan--Wigner transformation is seen as an obstacle for near-term devices and has led to the development of many alternatives in recent years.

Fermion-qubit mappings in the literature generally fall into two paradigms: ancilla-free mappings which use the minimum number of qubits to describe the fermionic algebra \cite{bravyi2002fermionic, seeley_bravyi-kitaev_2012, Jiang2020optimalfermionto, Vlasov_2022, miller2023bonsai, miller2024treespilation, harrison_sierpinski_2024, chiew2024ternarytreetransformationsequivalent} and local encodings which employ a number of ancilla qubits that scales linearly with the system size \cite{bravyi2002fermionic, verstraete2005mapping, chen2020exactarb, steudtner2019quantum, setia2019superfast, derby2021compact, chien2022optimizing, Algaba2024lowdepthsimulations, havlivcek2017operator}. 
As seen in Figure \ref{fig:different_mappings}, local encodings are designed to produce qubit Hamiltonians that are sums of local Pauli operators for fermionic systems such as the Fermi--Hubbard model \cite{hubbard1964}.
Meanwhile, ancilla-free mappings have been optimized to produce Pauli representations of fermionic hopping terms with logarithmically scaling support size through the Bravyi--Kitaev \cite{bravyi2002fermionic} and Ternary Tree transformations \cite{Vlasov_2022, Jiang2020optimalfermionto}.
However, in the near- and mid-term future for quantum computing, it is not necessarily the case that local encodings are always preferable to ancilla-free mappings \cite{cade_strategies_2020, nigmatullin2024experimentaldemonstrationbreakevencompact}, with qubit number itself being a scarce resource \cite{aspuru2005simulated} competing with the Pauli weights of the qubit Hamiltonian which serve as theoretical complexity parameters.

The fine line between ancilla-free mappings and local encodings presents remarkable opportunities for optimization, and there are numerous strategies to drive the costs of fermionic simulation lower avoiding the unwelcome scaling of ancillary qubits with the problem size.
In \cite{Chiew2023discoveringoptimal}, authors introduced a strategy to vary the fermionic labeling order in Jordan--Wigner transformations to reduce the total Pauli weight of the square lattice Fermi--Hubbard model;
an approach which has now expanded to Ternary Tree transformations \cite{parelladilme2024reducing}.
We also presented an analytic strategy to add two ancilla qubits to the Jordan--Wigner transformation, creating a hybrid mapping with a constant number of ancilla qubits. The addition of ancilla qubits allows cancellation of long strings of Pauli operators in the Hamiltonian terms, allowing a more dramatic reduction 
in Pauli weights over the ancilla-free version.

\noindent {\bf Our contribution: }In this work, we present two new strategies to further reduce the Pauli weight costs of fermion-qubit mappings in the absence of scaling qubit resources. We extend the method in \cite{Chiew2023discoveringoptimal} for optimal fermionic labeling to apply to a prominent class of ancilla-free mapping known as ``linear encodings" beyond Jordan--Wigner transformation, applying it in particular to the Bravyi--Kitaev, Ternary Tree and Parity Basis transformations. Through converting the task to a quadratic assignment problem, we also expand the scope of our strategy to fermionic systems with any hopping and interaction pattern, beyond just the square-lattice Fermi--Hubbard model.

Our second protocol generalises the two-ancilla qubit strategy from \cite{Chiew2023discoveringoptimal} to a method for incrementally adding 
any number of qubits to a Jordan--Wigner transformation. By also converting this task to a quadratic assignment problem, we expand the scope of these `constant-ancilla mappings' to Jordan--Wigner transformations for fermionic systems with any hopping and interaction model.
In the process, we demonstrate that with sufficient ancilla qubits, we can reduce the Pauli weight associated with the Jordan--Wigner transformation by as much as 67\%, improving upon even the best known ancilla-free transformations.
Furthermore, we determine optimal or near-optimal mappings for systems with several hundred fermions.
This work demonstrates that there is ample room for further improvement in fermionic simulation for near-term quantum computing, driving down the cost of quantum simulation avoiding significant overheads in quantum resources, through computationally-assisted searches and optimization.

\section{Related work}

In recent years, there have been various approaches to optimizing ancilla-free fermion-qubit mappings targeting different cost functions. In contrast to our approach in Section IIIA, in which we take a linear encoding and reduce total and average Pauli weight by optimizing over fermionic order, the space of linear encodings itself has been used as an optimization space to reduce costs such as the number of entangling gates in fermionic simulation \cite{wang2023evermore} and the locality of fermionic operators \cite{harrison_sierpinski_2024}. Generalisations of ternary tree transformations have optimized over tree shape to fit operators to qubit systems with limited connectivity \cite{miller2023bonsai} and to minimise the CNOT count in unitary coupled cluster ansatze \cite{miller2024treespilation}. Fermionic order optimizations have also further refined searches over tree shape to reduce the entanglement requirements in preparing target fermionic states \cite{parelladilme2024reducing}. Searches through Clifford conjugations of the Jordan--Wigner transformation to minimize the average Pauli weight of molecular Hamiltonians have explored a wider range of ancilla-free mappings \cite{yu2025cliffordcircuitbasedheuristic}.

The approach whereby one incrementally adds ancilla qubits to Jordan--Wigner transformations in order to reduce the cost was first explored in \cite{Chiew2023discoveringoptimal}. As far as we are aware, our work here represents the only subsequent development of this method. While local encodings have solved the problem of hopping terms with scaling Pauli weight, the outstanding requirement is a scaling number of ancilla qubits. The literature in this area has remained focused on reducing the linear scaling coefficient of ancilla qubits in local encodings, with the initial requirement being twice the number of modes \cite{verstraete2005mapping} reduced to smaller and smaller multiples \cite{chen2020exactarb,chen2022equivalence, Algaba2024lowdepthsimulations, steudtner2019quantum, derby2021compact, havlivcek2017operator, chien2022optimizing}, although always scaling linearly with system size.

\section{Background \& Notation} \label{sec:background}

In this section, we give a brief overview of fermionic systems and fermion-qubit mappings, defining the `linear encoding' class of ancilla-free mappings and establishing the conditions for existence of mappings with ancilla qubits. We introduce a variety of notational conventions that will be used throughout the paper, and adopt zero-indexing, i.e.\ $[j] = \{0,1,\dots,j-1\}$.

\subsection{Fermionic Systems vs Qubit Systems}
\label{sec:fermions}

Fermionic systems consist of \textit{modes} which represent lattice sites of quantized space in high energy and condensed matter physics or electronic orbitals in quantum chemistry. Modes can be occupied or unoccupied by single fermions, indistinguishable particles obeying the Pauli exclusion principle.

A mathematical description of fermionic statistics requires choosing an ordering $i \in [n]$ for the modes: while physically unimportant, the order enables the description of the $(2^n)$-dimensional Hilbert space of the system as  $\mathcal{H}_f^n = \text{span}\{\ketf{\vec{f}} \mid \vec{f} \in \mathds{Z}_2^n\}$, where the \textit{Fock basis state} $\ketf{\vec{f}}$ is the antisymmetrized wavefunction of an $n$--particle system with mode $i$ occupied if and only if $f_i=1$. The algebra of fermionic operators $B(\mathcal{H}_f^n)$ derives from the \textit{creation and annihilation operators} $a_i^{(\dagger)}$ for $i\in [n]$, which act on the Fock basis states via 
\begin{align}
    a_i^{(\dagger)} \ketf{\vec{f}} = 
    \begin{cases}
    (-1)^{\sum_{j<i} f_j} \ketf{(\vec{f} + \vec{e}_i) \bmod 2} & f_i = 1 \, (0) \\
    0 & f_i = 0 \, (1)\, ,
    \end{cases} \label{eqn:fockstates}
\end{align}
representing the addition or removal of a fermion in the $i$th mode from the overall system. Equation \eqref{eqn:fockstates} ensures the Pauli exclusion principle, which the canonical anticommutation relations express mathematically as 
\begin{align} \label{eqn:cars}
    \{a_i, a_j\} = 0 \, , \quad \{a_i^\dagger, a_j \} = \delta_{ij}\, ,
\end{align}
where $\{A,B\} = AB+BA$ for any two operators $A$ and $B$.

A \textit{fermionic Hamiltonian} is a Hermitian operator $H_f$ in $B(\mathcal{H}_f^n)$ that fully describes the fermionic system. It may contain \textit{number operators} $c_{ii} a_i^\dagger a_i$, the cost of single-particle tunnelling of fermions between modes via \textit{hopping terms} $c_{ij}a_i^\dagger a_j + c_{ij}^*a_j^\dagger a_i$, and the repulsive forces between fermions via \textit{interaction terms} $c^{ii}_{jj} a_i^\dagger a_i a^\dagger _j a_j$ for real coefficients $c_{ii},c_{jj}^{ii}$ and the possibly complex values $c_{ij}$.

In contrast, models for quantum computation assume access to \textit{qubits}, distinguishable spin-$\frac{1}{2}$ systems with state space $\mathcal{H}_2 = \text{span}\{\ket{0}, \ket{1}\}$. A quantum computer making use of an $m$-qubit register has access to the Hilbert space $\mathcal{H}_2^{\otimes m}$, which has the \textit{computational basis} $\{\ket{\vec{f}} \coloneqq \ket{f_0} \dots \ket{f_{n-1}} \mid \vec{f} \in \mathds{Z}_2^n\}$. Quantum computation consists of acting on qubit qubits with unitary operators, such as the Pauli matrices
\begin{align}
    X = \begin{pmatrix} 0 & 1 \\ 1 & 0 \end{pmatrix} \, , \quad Y = \begin{pmatrix} 0 & -i \\ i & 0 \end{pmatrix} \, ,\quad Z = \begin{pmatrix} 1 & 0 \\ 0 & -1 \end{pmatrix}\, .
\end{align}
We denote multi-qubit Pauli operators such as $X \otimes Y \otimes Z$ with the short-hand notation $X_0 Y_1 Z_2$.

\subsection{Fermion-qubit mappings and the Jordan--Wigner transformation} \label{sec:mappingsandjw}

Quantum simulation algorithms require a qubit Hamiltonian with the same spectral properties as the fermionic Hamiltonian of interest. This demands a unitary equivalence \cite{bravyi2002fermionic}, also known as a fermionic encoding \cite{chien2022optimizing} or \textit{fermion-qubit mapping}, from the operators and states of fermions to the language of qubits. Without exploiting symmetries of the Hamiltonian \cite{bravyi2017tapering, harrison2023reducingqubitrequirementjordanwigner, setia2020reducing}, there is a dimensional requirement of at least as many qubits ($m$) as there are modes ($n$). Formally, an \textit{ancilla-free fermion-qubit mapping} is an inner-product preserving map between $n$-mode and $n$-qubit space of the form $\mathfrak{m} : \mathcal{H}_f^n \rightarrow \mathcal{H}_2^{\otimes n}$. It is also useful to think of the induced transformation of fermionic operators to qubit operators:
\begin{align}
    \m{m} : \ketf{\psi} & \longmapsto  \ket{\psi}\, ,    & \ketf{\psi} &\in \mathcal{H}_f^n \, ;  \\
    \m{m} : a_i^{(\dagger)} & \longmapsto  \m{m} \cdot a_i^{(\dagger)} \cdot \m{m}^{\dagger}\, ,   &   i &\in [n]  \, .
\end{align}

The canonical example of an ancilla--free fermion-qubit mapping is the $n$-qubit \textit{Jordan--Wigner transformation} \cite{jordan:1928wi}, which maps each occupation number state $\ketf{\vec{f}}$ to the corresponding computational basis state:
\begin{align}
    \m{m}_\text{JW} :  \mathcal{H}_f^n \ni \ketf{\vec{f}} \longmapsto \ket{f_0} \ket{f_1} \dots \ket{f_{n-1}} \in \mathcal{H}_2^{\otimes n}\, .
\end{align}
The Jordan--Wigner transformation induces a transformation of the creation and annihilation operators,
\begin{align}
    \m{m}_\text{JW} : a_i^{(\dagger)} \longmapsto Z_0 Z_1 \dots Z_{i-1} \frac{X_i \mp Y_i}{2}\, , \label{eqn:jwas}
\end{align}
into sums of Pauli operators acting on up to $n$ qubits.

\subsection{Optimization problems for fermion-qubit mappings} \label{sec:problem}

The hopping and interaction terms in a fermionic system give a pattern of couplings between modes that can be described by a graph structure $G=(V,E)$ where \(V\) is a set of unordered vertices and \(E \subseteq \binom{V}{2}\) is a set of edges between vertices. 
In order to facilitate mapping our fermionic Hamiltonian to a qubit Hamiltonian, it is useful to fix an \textit{order} \(\sigma\colon V \to \left[\lvert V \rvert\right]\) for the vertices. Given \(\sigma\), we define
\begin{align}
    E_\sigma \coloneqq \left\{ \left(\sigma(v), \sigma(u)\right) \mid uv \in E, \sigma(v
    ) < \sigma(u)\right\}
\end{align}
to be the set of start and end positions of each edge with respect to the order $\sigma$.

A fermionic Hamiltonian corresponding to the graph $G$ takes the form
\begin{align}
    \label{eqn:probhamf}
    \begin{split}
    H_f &= \overbrace{\sum_{i \in [n]} c_i a_i^\dagger a_i}^{\text{number terms}} + 
    \overbrace{\sum_{\substack{(i,j) \in E_0}} c^{ii}_{jj} a_i^\dagger a_i a_j^\dagger a_j}^{\text{interaction terms}} \\
    & \quad + \underbrace{\sum_{\substack{(i, j) \in E_{0}}} (c_{ij} a_i^\dagger a_j + c_{ij}^* a_j^\dagger a_i}_{\text{hopping terms}}) 
    \end{split}
\end{align}
where \(E_0\) corresponds to an initial ordering of the fermionic labels.
Equation \eqref{eqn:probhamf} describes a large variety of quantum systems, including the Fermi--Hubbard model \cite{hubbard1964} as well as quantum chemical Hamiltonians \cite{mcclean2014exploiting}.

To generate a qubit representation of the fermionic Hamiltonian, consider applying a fermion-qubit mapping $\m{m}$ to obtain
\begin{align}
    H_{q}&= \sum_{i \in [n]} c_i A_i^\dagger A_i + \sum_{(i,j) \in E_0} c^{ii}_{jj} A_i^\dagger A_i A_j^\dagger A_j  \label{eqn:outputham} \\ 
    &\quad  + \sum_{(i,j) \in E_0} (c_{ij} A_i^\dagger A_j + c_{ij}^*A_j^\dagger A_i)\, , \nonumber
\end{align}
where $A_i^{(\dagger)} = \m{m} \cdot a_i^{(\dagger)} \cdot \m{m}^\dagger$. We can assess the quality of the fermion-qubit mapping $\m{m}$ in simulating $H_f$ in terms of the following two cost functions of the qubit Hamiltonian:
\begin{enumerate}
    \item The \textit{total Pauli weight}, which is equal to the sum of the number of qubits upon which each term in $H_q$ acts. The total Pauli weight could correspond to the number of single-qubit measurements required to measure each term of the Hamiltonian once, relating, for example, to ground-state estimation as part of a VQE procedure.
    \item The \textit{maximum Pauli weight}, which is equal to the maximum number of qubits upon which any one term in $H_q$ acts. The maximum Pauli weight corresponds to the non-locality of the Hamiltonian, a time complexity parameter of simulation algorithms such as Trotterization.
\end{enumerate}

\begin{figure}
    \centering
    \includegraphics[width=\linewidth]{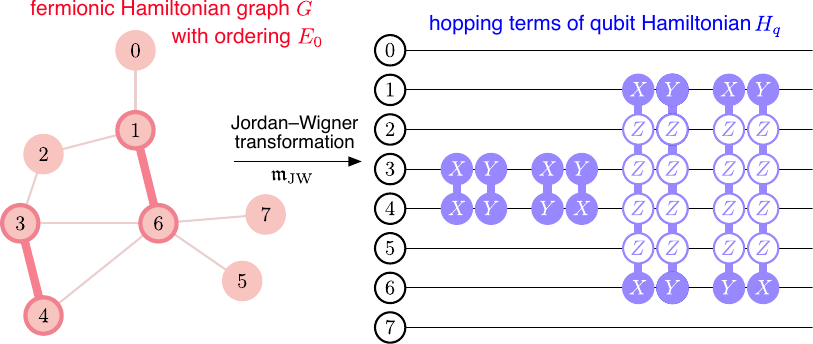}
    \caption{The Jordan--Wigner transformation of fermionic hopping terms into qubit operators, which uses no ancilla qubits.}
    \label{fig:hoppingterms}
\end{figure}

For example, under the Jordan--Wigner transformation, the number, hopping and interaction terms in \eqref{eqn:probhamf} map respectively to the following qubit operators:
\begin{align}
     \m{m}_{\text{JW}} : {a_i^\dagger a_i} & \longmapsto \frac{1}{2}(\mathds{1}^{\otimes n} - Z_i) \\
    \label{eqn:jwhopping}
    {c_{ij} a_i^\dagger a_j + c_{ij}^* a_j^\dagger a_i}  & \longmapsto 
    \big( \text{Re}(c_{ij})(X_i X_j + Y_i Y_j)  \\
    &\quad \,  +  \text{Im}(c_{ij}) (X_i Y_j + Y_i X_j) \big) 
    Z_{[i+1,j-1]}
    \nonumber \\
    {a_i^\dagger a_i a_j^\dagger a_j} &\longmapsto \frac{1}{4}(\mathds{1}^{\otimes n} - Z_i)(\mathds{1}^{\otimes n} - Z_j)\, .
\end{align}
Each hopping term maps to a sum of four Pauli operators, each with weight $|i{-}j|{+}1$, as in Figure \ref{fig:hoppingterms}. The number and interaction terms map to local sums of Pauli operators with weights 1 and 2, respectively, regardless of the total number of modes $n$.
The presence of hopping terms means that overall, the Jordan--Wigner transformation produces terms of weights up to $n$ in simulation tasks for $H_f$. In contrast, other ancilla-free mappings such as the Bravyi--Kitaev \cite{bravyi2002fermionic} and Ternary Tree transformations \cite{Vlasov_2022, Jiang2020optimalfermionto} produce qubit representations of the hopping terms with weights up to ${\sim}\log_2 n$ and ${\sim} \log_3(2n)$. While this is an advantage over the Jordan--Wigner transformation in terms of total and maximum Pauli weight, we should note that the number and interaction terms do not transform into local qubit operators under these other mappings.

\subsection{Majorana operators and linear encodings}\label{sec:linear}

In this paper, we will primarily work with the fermionic operators known as \textit{Majorana operators} \(\gamma_{i}, \overline{\gamma}_{i}\), which are defined for $i \in [n]$ as 
\begin{align}
    \gamma_{i} &\coloneqq a_i + a_i^\dagger\, , & \overline{\gamma}_{i} &\coloneqq -\iu \left(a_i - a_i^\dagger\right)\, .
\end{align}
In the Majorana basis, for $i,j \in [n]$ the canonical anticommutation relations of \eqref{eqn:cars} translate to
\begin{align}
    \{\gamma_{i}, \overline{\gamma}_{j}\} = 2 \delta_{ij} \, ,  \quad  \{\gamma_{i}, \gamma_{j} \} = \{ \overline{\gamma}_i, \overline{\gamma}_j\} = 0\, ; \\
    \gamma_{i}^\dagger = \gamma_{i}\, , \quad \overline{\gamma}_i^\dagger = \overline{\gamma}_i \, .
\end{align}

\textit{Linear encodings} are a prominent class of fermion-qubit mappings that act as binary codes of the Fock basis, yielding the transformation from fermionic to qubit states
\begin{align}
    \m{m}_{U} :   \mathcal{H}_f^n \ni \ketf{\vec{f}} \longmapsto \ket{U \vec{f}} \in \mathcal{H}_2^{\otimes n}\label{eqn:linearmapping}
\end{align}
for an invertible binary matrix $U \in \text{GL}_n(\mathbb{F}_2)$. For example, the $n$-qubit Jordan--Wigner transformation is trivially the linear encoding with $U = \mathds{1}$, the $n{\times}n$ identity matrix.

Fermionic operators transform predictably under linear encodings \cite{chiew2024ternarytreetransformationsequivalent, harrison_sierpinski_2024}. The Majorana representations of the mapping in \eqref{eqn:linearmapping} are
\begin{align}\label{eq:majorana_rep}
    \m{m}_U : \begin{cases}
        \gamma_i \longmapsto X_{U(i)} Z_{P(i)} \\
        \overline{\gamma}_i \longmapsto X_{U(i)} Z_{R(i)} \, .
    \end{cases}
\end{align}
Here, the expression $U(i)$ denotes the row indices of nonzero elements in the $i$th column of $U$. If the expression $F(i)$ denotes the column indices of nonzero elements in the $i$th row of $U^{-1}$, then $P(i) = F(0) \, \triangle \, F(1) \, \triangle \, \dots \, \triangle \, F(i-1)$ and $R(i) = P(i) \, \triangle \, F(i)$, where $\, \triangle \,$ denotes the symmetric difference of two sets: $A \, \triangle \, B = (A \cup B ) \backslash (A \cap B)$. 

In addition to the Jordan--Wigner transformation, there are three other well-known linear encodings. The \textit{Parity Basis transformation} takes the form
\begin{align}
    \m{m}_\text{PB} : \ketf{\vec{f}} \longmapsto \ket{L \vec{f}}\, ,
\end{align}
where $L$ is the lower-triangular matrix of 1s, including diagonal entries. The \textit{Bravyi--Kitaev transformation} \cite{bravyi2002fermionic, seeley_bravyi-kitaev_2012} has a canonical definition when $n=2^k$ is a power of 2, with
\begin{align}
    \m{m}_\text{BK} : \ketf{\vec{f}} &\longmapsto \ket{U_\text{BK} \vec{f}}\, ,
\end{align}
where $U_\text{BK} = U_\text{BK}^{(k)}$, the $k$th matrix in the sequence
\begin{align}\label{eq:bk}
    U_\text{BK}^{(0)} = 1 \, , \quad U_\text{BK}^{(i+1)} = \begin{pmatrix}
    U_\text{BK}^{(i)} & \scalebox{0.5}{$\begin{matrix} 0 & 0 & \dots & 0 \\ 0 & 0 & \dots & 0 \\ \vdots & \vdots & \ddots & \vdots \\ 0 & 0 & \dots & 0 \end{matrix}$} \\
    \scalebox{0.5}{$\begin{matrix} 0 & 0 & \dots & 0 \\ 0 & 0 & \dots & 0 \\ \vdots & \vdots & \ddots & \vdots \\ 1 & 1 & \dots & 1 \end{matrix}$} & U_\text{BK}^{(i)}
\end{pmatrix} \, .
\end{align}
The \textit{Ternary Tree transformation} \cite{Jiang2020optimalfermionto, Vlasov_2022} has a canonical definition when $n = (3^k+1)/2$ for some integer $k$. The original presentations specified the Majorana representations via a tree graph without defining the encoded Fock states, but it has become apparent \cite{harrison_sierpinski_2024, chiew2024ternarytreetransformationsequivalent} that the Ternary Tree transformation is effectively the linear encoding with definition
\begin{align}
    \m{m}_\text{TT} : \ketf{\vec{f}} \longmapsto \ket{U_\text{TT} \vec{f}}\, ,
\end{align}
where  $U_\text{TT} = U_\text{TT}^{(k)}$, the  $k$th matrix in the sequence
\begin{align} \label{eqn:tt}
    U_\text{TT}^{(0)} = 1 \, , \quad U_\text{TT}^{(i+1)} =\begin{pmatrix}
    U_\text{TT}^{(i)} & 0 & 0 & 0 \\
    1 & 1 & 1 & 0 \\
    0 & 0 & (U_\text{TT}^{(i)})^{\top'} & 0 \\
    0 & 0 & 0 & U_\text{TT}^{(i)}
\end{pmatrix}\, ,
\end{align}
and where $U^{\top '}$ denotes the transpose of $U$ along the main diagonal, with coordinates $(U^{\top'})_{i,j} = U_{n-j,n-i}$.

While the Bravyi--Kitaev and Ternary Tree transformations are only formally defined for restrictive numbers of fermions, there has been an effort to interpolate these mappings to any \(n\). The Bravyi--Kitaev and Ternary Tree transformations correspond to the 
Fenwick \cite{havlivcek2017operator}
and Sierpiński tree data structures \cite{harrison_sierpinski_2024}, respectively. These mappings can be extended to systems of arbitrary size by instead using a pruned version of the next largest tree of each type. This is the convention that we utilize throughout this paper.

\subsection{Using ancilla qubits in fermion-qubit mappings}\label{sec:local}

\begin{figure}
    \centering
    \includegraphics[width=\linewidth]{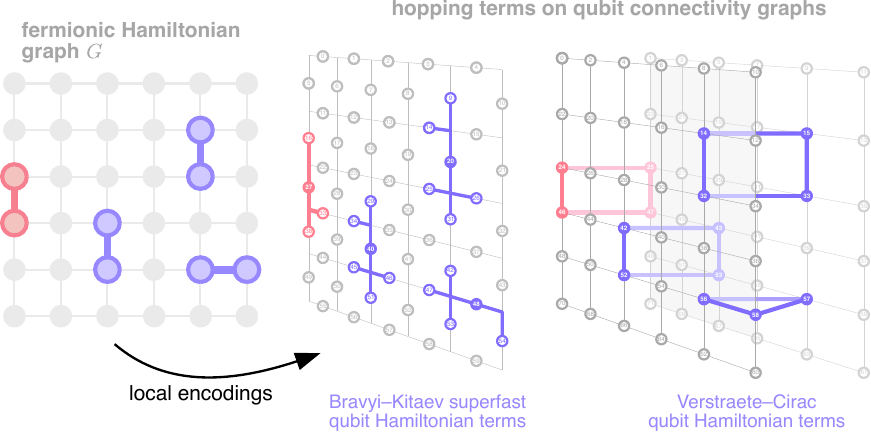}
    \caption{Local encodings ensure the Pauli weights of hopping terms are bounded by a constant $\mathcal{O}(1)$ by preserving the locality of the Hamiltonian graph. This property comes at the cost of using $\mathcal{O}(n)$ ancilla qubits.}
    \label{fig:localencodings}
\end{figure}

\begin{figure}
    \centering    \includegraphics[width=\linewidth]{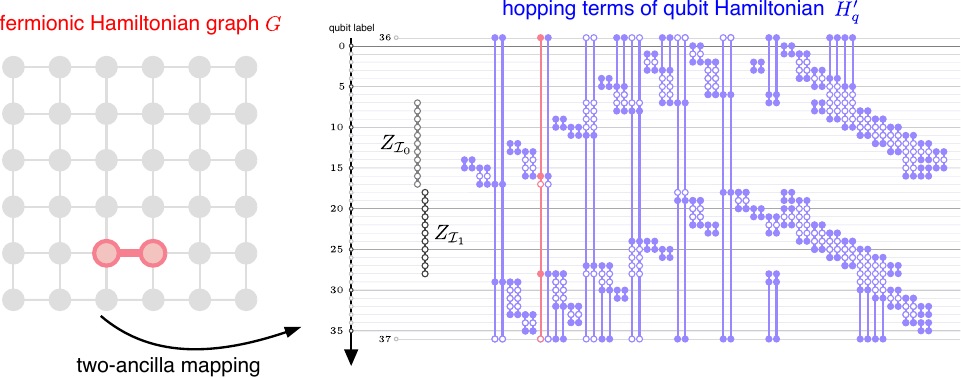}
    \caption{Two-ancilla mapping producing a lower maximum and total Pauli weight than any ancilla-free mapping while using only $\mathcal{O}(1)$ ancilla qubits.}
    \label{fig:6x6aug2}
\end{figure}

The linear encodings of Section \ref{sec:linear} are ancilla-free mappings, converting between fermionic systems and quantum computers with the same numbers of modes and qubits. In general, a fermion-qubit mappings is any inner-product preserving map $\m{m} : \mathcal{H}_f \rightarrow \mathcal{H} < \mathcal{H}_2^{\otimes m}$ between an $n$-mode fermionic system and a $(2^n)$-dimensional subspace of $m$-qubit qubit register, where $m\geq n$.
In contrast to ancilla-free mappings between $n$ modes and $n$ qubits, the larger number of qubits available can enable a reduction in Pauli weights of operators of interest by virtue of the restriction to the subspace $\mathcal{H}$. Theorem \ref{thm:hequiv} provides a useful lens through which to view fermion-qubit mappings with ancilla qubits.

\begin{theorem} \label{thm:hequiv} \textit{(Adapted from \cite{chien2022optimizing}.)} 
Let $H_f$ be an $n$-mode fermionic Hamiltonian and let $H_q$ be an $m$-qubit Hamiltonian, where $m\geq n$. The following statements are equivalent:
\begin{enumerate}
    \item \label{thm:p1} There is a fermion-qubit mapping $\m{m} : \mathcal{H}_f^n \rightarrow \mathcal{H} \leq \mathcal{H}_2^{\otimes m}$ between the two Hamiltonians, i.e.\ a unitary equivalence
        \begin{align}
        H_q = \m{m} \cdot H_f \cdot \m{m}^\dagger \, ,
        \end{align}
        where the $(2^n)$-dimensional Hilbert space $\mathcal{H}$ is the space stabilized by any qubit operator in the kernel of $\m{m}^\dagger$. That is, if the qubit operator $A \in B(\mathcal{H}_2^{\otimes m})$ is an image of the fermionic identity in the sense that $\m{m}^\dagger (A) = 1$, then $A \ket{\psi} = \ket{\psi}$ for all $\ket{\psi} \in \mathcal{H}$. 
    \item \label{thm:p2} Let the fermionic Hamiltonian have the decomposition into the sum of products of Majorana operators
    \begin{align}
        H_f = \sum_{\mathclap{\mathcal{I}, \mathcal{J} \subset [n]}} (c_{\mathcal{I}, \mathcal{J}}) \gamma_{\mathcal{I}, \mathcal{J}} \, , \, \text{ where }  \gamma_{\mathcal{I},\mathcal{J}} = \prod_{\substack{i \in \mathcal{I} \\ j \in \mathcal{J}}} \gamma_{i} \overline{\gamma}_j\, .
    \end{align}
    Then, the qubit Hamiltonian has a similar decomposition into Pauli operators, i.e.
    \begin{align}
        H_q = \sum_{\mathclap{\mathcal{I}, \mathcal{J} \subset [n]}} (c_{\mathcal{I},\mathcal{J}}) P_{\mathcal{I}, \mathcal{J}} \quad \text{for Paulis } P_{\mathcal{I}, \mathcal{J}}\, . 
    \end{align}
    The Hamiltonian terms have the same Hermiticity,
    \begin{align}
        (\gamma_{\mathcal{I}, \mathcal{J}})^\dagger  = \pm \gamma_{\mathcal{I}, \mathcal{J}} \iff (P_{\mathcal{I}, \mathcal{J}})^\dagger = \pm P_{\mathcal{I}, \mathcal{J}}\, ,
    \end{align}
    and the same commutation relations:
    \begin{align}
        \{\gamma_{\mathcal{I}, \mathcal{J}}, \gamma_{\mathcal{I}', \mathcal{J}'}\} = 0 &\iff \{P_{\mathcal{I}, \mathcal{J}}, P_{\mathcal{I}', \mathcal{J}'}\} \, ,\\
        [\gamma_{\mathcal{I}, \mathcal{J}}, \gamma_{\mathcal{I}', \mathcal{J}'}] = 0 &\iff [P_{\mathcal{I}, \mathcal{J}}, P_{\mathcal{I}', \mathcal{J}'}] \, .
    \end{align}
\end{enumerate}
\end{theorem}

Figure \ref{fig:localencodings} displays several terms from the Bravyi--Kitaev superfast \cite{bravyi2002fermionic} and Verstraete--Cirac transformations \cite{verstraete2005mapping}, examples of local encodings that produce $\mathcal{O}(1)$-weight hopping terms using $n{+}\mathcal{O}(n)$ qubits.
The Bravyi--Kitaev superfast method constructs an $m$-qubit Hamiltonian where $m$ is the number of edges in the Hamiltonian graph,
and carefully chooses Pauli operators obeying statement \ref{thm:p2} of Theorem \ref{thm:hequiv} that act only locally with weight $\mathcal{O}(1)$ upon each edge of the graph. The fermion-qubit mapping thus gives a qubit Hamiltonian with significant reduction upon the $\mathcal{O}(n)$-weight hopping terms of the Jordan--Wigner transformation from Section \ref{sec:problem}.

While local encodings promise far simpler qubit Hamiltonians than ancilla-free mappings, these results are only beginning to bear fruit on near-term devices \cite{cade_strategies_2020, nigmatullin2024experimentaldemonstrationbreakevencompact}. The requirement of $\mathcal{O}(n)$ ancilla qubits can be a heavy price to pay for Pauli weight reduction when logical qubits themselves are a scarce resource.

In Section \ref{sec:constantancillas}, we generalize a method from \cite{Chiew2023discoveringoptimal} into a strategy for incrementally adding ancilla qubits to Jordan--Wigner transformations, resulting in \textit{constant-ancilla mappings}. Figure \ref{fig:6x6aug2} gives a two-ancilla mapping as an example.

\subsection{Order Optimization}
\begin{figure}
    \centering
    \includegraphics[width=\linewidth]{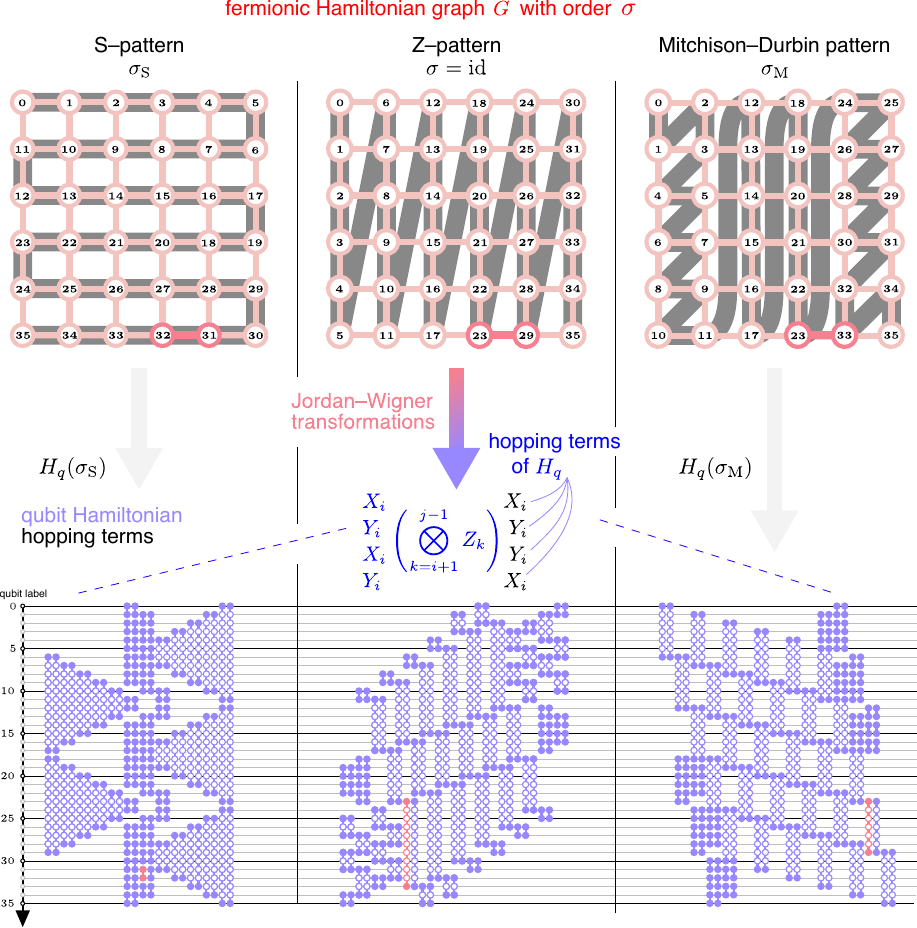}
    \caption{Different qubit Hamiltonians resulting applying the Jordan--Wigner transformation to the $6{\times}6$ Fermi--Hubbard model with different fermionic labelings.}
    \label{fig:signatures}
\end{figure}

Regardless of the linear encoding at hand, the fermionic order yields qubit Hamiltonians with different total and maximum Pauli weights, despite being a physical symmetry of the fermionic system in question. As Figure \ref{fig:signatures} illustrates, our approach in Section \ref{sec:1sum} exploits the freedom in fermionic order to minimize total or maximum Pauli weight. 

The problem of ordering the vertices of a graph to minimize some given cost is ubiquitous, exemplified by the well studied Quadratic Assignment Problem \cite{LOIOLA2007657}. For a given graph \(G=(V,E)\) and distance matrix \(\mathcal{D},\) the unweighted form of the quadratic assignment problem can be written as
\begin{align*}
    \min_{\sigma} \sum_{uv \in E} \mathcal{D}_{\sigma(u) \sigma(v)}.
\end{align*}

There are a variety of search based methods for solving Quadratic Assignment in its most general form\cite{SILVA20211066}. However, for certain choices of distance matrix \(\mathcal{D},\) Quadratic Assignment admits efficient heuristic approaches. For example, Minimum Linear Arrangement (MLA) is a special case of Quadratic Assignment with \(\mathcal{D}_{ij} = \lvert i - j\rvert\). One can find high quality solutions for MLA very quickly by utilizing multilevel methods, even for large graph inputs \cite{safro_order, Ibrahim:2022wju}.

\begin{figure*}
    \centering
    \includegraphics[width=\linewidth]{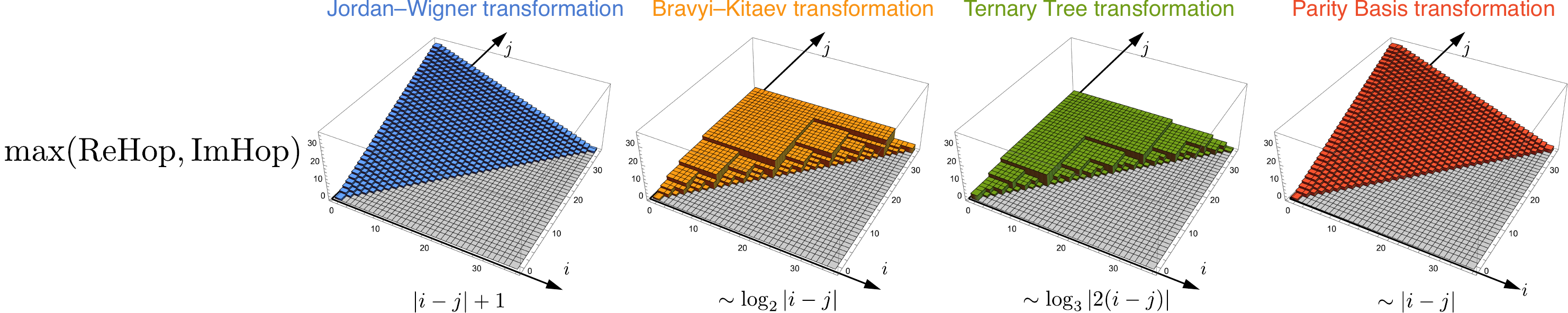}
    \caption{Maximum Pauli weight surface $\mathcal{D} = \max(\text{ReHop}, \text{ImHop})$ of the hopping terms corresponding to the edge $(i,j)$ in a fermionic Hamiltonian graph via the Jordan--Wigner, Bravyi--Kitaev, Ternary Tree and Parity Basis transformations. The Jordan--Wigner transformation has a closed-form cost function, and hence makes a good target for optimization routines.}
    \label{fig:surfaces}
\end{figure*}

\section{Our Methodology} \label{sec:methodology}

We consider two approaches to optimization over fermion-qubit mappings to reduce the maximum and total Pauli weights of Hamiltonians of the form in \eqref{eqn:outputham}. The first approach, in Section \ref{sec:1sum}, fixes an arbitrary linear encoding $\m{m}_U$ and optimizes over the fermionic order $\sigma$ to minimize the Pauli weight of the qubit Hamiltonian.
This generalizes the method from \cite{Chiew2023discoveringoptimal} which previously only applied to the Jordan--Wigner transformation.

The second approach, in Section \ref{sec:constantancillas}, introduces a new strategy to incrementally add ancilla qubits to the Jordan--Wigner transformation in order to produce a qubit Hamiltonian with lower maximum and total Pauli weights than the mappings with optimized orderings from Section \ref{sec:1sum}. This method generalises the approach in \cite{Chiew2023discoveringoptimal} to add more than two qubits, and applies to arbitrary fermionic Hamiltonian graphs rather than just the square lattice.

The following fermionic operator identities are helpful:
\begin{align}
        &a_i^{(\dagger)} = \frac{1}{2} (\gamma_{i} \pm \iu \overline{\gamma}_{i}) \, , \quad  
         a_i^\dagger a_i = \frac{1}{2}(1+\iu \gamma_{i}\overline{\gamma}_{i}) \\
        \label{eqn:majoranahopping}
        &c_{ij}a_i^\dagger a_j + c_{ij}^* a_j^\dagger a_i =  \text{Re}(c_{ij}) (\gamma_{i} \gamma_j + \overline{\gamma}_i \overline{\gamma}_{j})\\
        &\qquad \qquad \qquad \qquad \quad + \iu \, \text{Im}(c_{ij})(\gamma_{i}\overline{\gamma}_{j} - \overline{\gamma}_{i} \gamma_{j}) \, .\nonumber 
\end{align}

\subsection{Fermionic Order as Quadratic Assignment}\label{sec:1sum}

While a fermion-qubit mapping may produce the Hamiltonian $H_q$ in \eqref{eqn:outputham} from the fermionic Hamiltonian $H_f$ in \eqref{eqn:probhamf},
as outlined in Section \ref{sec:fermions}, the order of the fermionic mode is arbitrary and does not affect the underlying dynamics of the model \cite{Chiew2023discoveringoptimal}.
Thus,
we have the freedom to relabel the fermionic modes and consider qubit Hamiltonians of the form
\begin{align} \label{eqn:qhamsigma}
    H_q(\sigma) &= \sum_{i \in [n]} c_i A_i^\dagger A_i + \sum_{(i,j) \in E_{\sigma}} c_{jj}^{ii} A_i^\dagger A_i A_j^\dagger A_j \\
    & \quad + \sum_{(i,j) \in E_{\sigma}} (c_{ij}A_i^\dagger A_j + c_{ij}^* A_j^\dagger A_i)\, .
\end{align}
for any order $\sigma : V \rightarrow [n]$. For example, Figure \ref{fig:signatures} demonstrates the hopping terms arising from the Jordan--Wigner transformation of the Fermi--Hubbard model using three distinct fermionic labelings.

Let $\m{m}_U$ be a linear encoding for some $n{\times}n$ invertible binary matrix $U$. Using \eqref{eq:majorana_rep} and \eqref{eqn:majoranahopping}, it is straightforward to evaluate the representations of hopping terms under $\m{m}_U$:
\begin{align} \label{eqn:qubithopping}
    & \quad  \frac{1}{2}   \left( \Im(c_{ij})(X_{U_{i \triangle j}} Z_{P_{i \triangle j}} + X_{U_{i \triangle j}} Z_{R_{i \triangle j}} ) \right. \\
    & \qquad \quad  \left. - \Re(c_{ij}) (X_{U_{i \triangle j}} Z_{P_i \triangle R_j} - X_{U_{i \triangle j}} Z_{R_i \triangle  P_j})\right) \nonumber
\end{align}
where we have used the shorthand $U_i$ and  $U_{i\triangle j}$ to denote  $U(i)$ and $U(i) \, \triangle \, U(j)$, respectively, and likewise for the $F$ and $R$ sets. We have also used the rules that $X_{U(i)} Z_{P(i)} = Z_{P(i)} X_{U(i)}$ and $X_{U(i)} Z_{R(i)} = - Z_{R(i)} X_{U(i)}$ from Lemma 7.2 of \cite{harrison_sierpinski_2024}. Using \eqref{eqn:majoranahopping}, the interaction terms transform to
\begin{align} \label{eqn:qubitinteraction}
    \frac{1}{4}(\mathds{1} - (Z_{P_i \triangle R_i} + Z_{P_j \triangle R_j}) + (Z_{P_i \triangle R_i} Z_{P_j \triangle R_j}))\, .
\end{align}

Given an invertible binary matrix \(U \in \text{GL}_n(\mathbb{F}_2)\), define the binary matrices
\begin{align*}
    F = U^{-1}\, , \quad   
    P = LF \, , \quad 
    R = P + F 
\end{align*}
where {\(L_{ij} = 1\) if \(i < j\)} and \(0\) otherwise. Each column of \(U\) and each row of \(F,P,R\) then correspond to the set families of the same name in \eqref{eq:majorana_rep}, e.g. {\(U_{\ast j} = U(j)\) and \(F_{i} = F(i)\)}.

Using \eqref{eqn:qubithopping} and \eqref{eqn:qubitinteraction}, we may use the matrix definitions of $F$, $P$ and $R$ to define a matrix \(\mathcal{D} \in \mathbb{N}^{n \times n}\) representing the total ($\oplus = +$) or maximum ($\oplus = \max$) Pauli weight of \(H_q(\sigma)\) consisting of the elementwise \(\oplus\)-sum of the weight of {number} terms, hopping terms with a nontrivial real part, hopping terms with a nontrivial complex part, and interaction terms:
\begin{align}
    \begin{split}
    \text{Num}_{i} &\coloneqq \lVert F_i\rVert_0\\
    \text{ReHop}_{ij} &\coloneqq \lVert \max(P_i + R_j,  U_{\ast i} + U_{\ast j})\rVert_0 \oplus \phantom{123}\\
    &\phantom{\coloneqq}\qquad\lVert \max(R_i + P_j,  U_{\ast i} + U_{\ast j})\rVert_0\\
    \text{ImHop}_{ij} &\coloneqq \lVert \max(P_i + P_j,  U_{\ast i} + U_{\ast j})\rVert_0 \oplus \phantom{123}\\
    &\phantom{\coloneqq}\qquad\lVert \max(R_i + R_j,  U_{\ast i} + U_{\ast j})\rVert_0\\
    \text{Inter}_{ij} &\coloneqq\lVert F_i\rVert_0 \oplus \lVert F_j\rVert_0 \oplus  \lVert F_i + F_j\rVert_0
    \end{split}
\end{align}
A particular model may include any or all components:
\begin{align*}
    \mathcal{D}_{ij} = \text{ReHop}_{ij} \oplus \text{ImHop}_{ij} \oplus \text{Inter}_{ij} \, , \quad \mathcal{D}_{ii} = \text{Num}_i\, . 
\end{align*}
We assume that the hopping terms of the problem Hamiltonian $H_f$ have either (1) purely real or imaginary coefficients, so that one of ImHop and ReHop is zero, or (2) complex coefficients in which case both ImHop and ReHop are necessary to calculate Pauli weights of hopping terms.
In cases that combine these two scenarios, the expression $\mathcal{D}_{ij}$ acts as an upper bound.
The total or maximum Pauli weight is thus
\begin{align}
    \left(
    \bigoplus_{(i,j) \in E_\sigma } \mathcal{D}_{ij}
    \right)
    \oplus
    \left(
    \bigoplus_{i \in V_\sigma} \mathcal{D}_{ii}
    \right)\, .
\end{align}

 We give examples for $\mathcal{D}$ with $\oplus = \max$ in Figure \ref{fig:surfaces} for the Jordan--Wigner, Bravyi--Kitaev, Ternary Tree and Parity Basis transformations. This yields the optimization problem
\begin{align}\label{eq:order_opt}
    \min_{\sigma} \left( \bigoplus_{(i,j) \in E_\sigma } \mathcal{D}_{ij} \right) \oplus \left( 
    \bigoplus_{i \in V_\sigma} \mathcal{D}_{ii} \right) \, .
\end{align}
{This is the problem that we solve in Section \ref{sec:1sumres}.}
{Minimizing the total Pauli weight corresponds to quadratic assignment, while minimizing the maximum Pauli weight relates to the bandwidth problem for Minimum Linear Arrangement.}

\subsection{Incremental addition of ancilla qubits} \label{sec:constantancillas}

In this section we generalize the approach in \cite{Chiew2023discoveringoptimal} to a procedure for incrementally adding ancilla qubits to Jordan--Wigner transformations.
We first show how a mapping using a single extra ancilla qubit can produce a Hamiltonian consisting of terms with lower maximum and total Pauli weights; the  incremental procedure proceeds inductively.
Theorem \ref{thm:hequiv} is the theoretical bedrock of our method;
we make frequent use of statement \ref{thm:p2} to ensure the validity of the modified qubit Hamiltonian as a representation of the target fermionic system.

Begin with an $n$-mode fermionic Hamiltonian $H_f$ of the form in \eqref{eqn:probhamf} and the resulting qubit Hamiltonian generated by the $n$-qubit ancilla-free Jordan--Wigner transformation,
\begin{align}
    \label{eqn:jwancil} 
    H_q(\sigma) = \sum_{(i,j) \in E_{\sigma}}  \underbrace{( \dots )_{ij}
    Z_{[i+1,j-1]}
    }_{\text{hopping term}} \, + \, \dots \, ,
\end{align}
where {$\sigma: V \rightarrow [n]$ is an optimized fermionic label order for total or maximum Pauli weight from Section \ref{sec:1sum}, depending on the cost function of choice.} Unlike in \eqref{eqn:jwhopping}, Equation \eqref{eqn:jwancil} conceals the linear combination of the two-qubit operators $X_iX_j$, $Y_iY_j$, $X_i Y_j$ and $Y_i X_j$ as $(...)_{ij}$  for brevity. The unwritten terms in \eqref{eqn:jwancil} are the the number and interaction terms, which are local under the Jordan--Wigner transformation. As per statement \ref{thm:p2}, the terms of $H_q(\sigma)$ have the same Hermiticity and commutation relations relative to the terms in $H_f$.

\begin{figure}
	\centering
    \includegraphics[width=\linewidth]{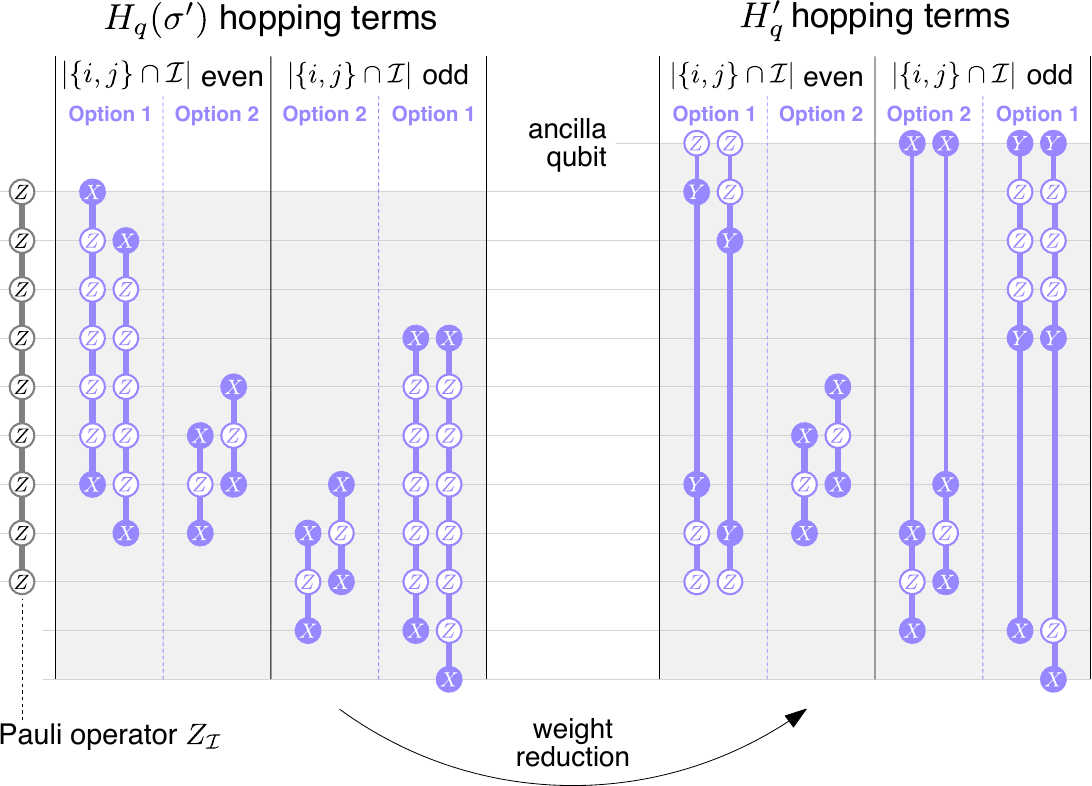}
	\caption{One iteration of the method for incrementally adding ancilla qubits to Jordan--Wigner transformations and the corresponding reduction in Pauli weights of a qubit Hamiltonian. The ancilla comes with the user's choice of Pauli string $Z_{\mathcal{I}}$: in constructing the Hamiltonian $H_q'$, which is equivalent to $H_q(\sigma')$, the user has the choice to modify each term of the original Hamiltonian by $Z_\mathcal{I}$ or leave it unchanged up to local action on the ancilla, leading to a reduction in Pauli weights.}
	\label{fig:kappas}
\end{figure}

We wish to construct an $(n{+}1)$-qubit Hamiltonian $H_q'$ that also satisfies statement \ref{thm:p2}, while reducing the Pauli weights of some of the hopping terms. 
Define $H_q'$ to be an $(n{+}1)$-qubit Hamiltonian with identical number and interaction terms to $H_q(\sigma)$, but with each hopping term modified via
\begin{align}
     (\cdots)_{ij}
     Z_{[i+1,j-1]}
     &\mapsto (\cdots)_{ij}
     Z_{[i+1,j-1]}
     \kappa_{ij}.
\end{align}
The Pauli operator \(\kappa_{ij}\) may take one of two forms for each individual term, chosen to minimize the Pauli weight of the resulting term. These options are 
\begin{align}
\begin{split}\label{eq:kappas}
    \text{(Option 1)}\qquad\kappa_{ij} &\coloneqq \begin{cases}
        Z_\mathcal{I} \otimes Y_n & \lvert \{i,j\} \cap \mathcal{I}\rvert = 1\\
        Z_\mathcal{I} \otimes Z_n & \text{otherwise}
    \end{cases} \\
    \text{(Option 2)} \qquad \kappa_{ij} &\coloneqq \begin{cases}
         \mathds{1}^{\otimes n} \otimes X_n & \lvert \{i,j\} \cap \mathcal{I}\rvert = 1\\
        \mathds{1}^{\otimes n} \otimes \mathds{1}_n & \text{otherwise}
    \end{cases}
\end{split}
\end{align}
Options are independent for each hopping term of the Hamiltonian in order to minimize Pauli weight -- one term utilizing Option 1 does not exclude another from utilizing Option 2.

In Option 1, the operator \(\kappa_{ij}\) is the product of $Z_{\mathcal{I}},$ a string of Pauli $Z$-gates acting on some subset $\mathcal{I} \subset [n]$ of qubits, and a single-qubit Pauli operator acting on the ancilla. {In Option 2, the operator $\kappa_{ij}$ is simply a Pauli operator acting on the ancilla qubit.} The string $Z_{\mathcal{I}}$ serves the purpose of reducing the Pauli weight of the hopping terms, while the  ancilla gates preserve the commutation relations of the terms of $H_q(\sigma)$ in the terms of $H_q'$ as per the requirements of statement \ref{thm:p2}.  Figure \ref{fig:kappas} illustrates the process of generating the modified Hamiltonian $H_q'$ given the terms of $H_q(\sigma)$.

The question is then, how much does \(\kappa_{ij}\) affect the Pauli weight of a Jordan--Wigner hopping term? For Option 2, this is fairly trivial: the weight of an $(i,j)$ hopping term in the new Hamiltonian $H_q^\prime$ increases by $|\{i,j\} \cap \mathcal{I}| \bmod 2$, corresponding to the case where a nontrivial gate is placed on the ancilla.

For Option 1, we may ignore any \(Z\)s applied to the endpoint qubits $i$ and $j$ of the $(i,j)$ hopping term since \(XZ \propto Y\) and \(YZ \propto X.\) The Pauli weight is then reduced by the number of indices in \(\mathcal{I}\) falling strictly between \(i\) and \(j,\) but increased by the number falling outside that range, plus one for the ancilla gate. {The weight of the $(i,j)$ hopping term in $H_q'$ increases by the possibly negative quantity}
\begin{align}
    d_{ij}(\mathcal{I}) &\coloneqq \lvert \mathcal{I} \cap ([n] \setminus [i, j])\rvert - \lvert \mathcal{I} \cap [i+1, j-1]\rvert + 1\, .
\end{align}

Overall, this yields an optimization problem to find the {choice of qubits $\mathcal{I}$ that minimizes the increase in Pauli weights of the terms in $H_q'$. The solution
minimizes}
\begin{align}
\begin{split}
\min_{\mathcal{I}\subset [n]}\quad &\sum_{(i,j) \in E_\sigma}\alpha_{ij}\min\Big(\lvert \{i,j\} \cap \mathcal{I}\rvert \bmod 2, d_{ij}(\mathcal{I})\Big)
\end{split}
\end{align}
where the weighting \begin{align*}
    \alpha_{ij} \coloneqq \begin{cases}
        4 & \text{Im}(c_{ij})\neq 0 \neq \text{Re}(c_{ij}) \\
        0 & \text{Im}(c_{ij}) = 0 = \text{Re}(c_{ij})\\
        2 & \text{otherwise}
    \end{cases}
\end{align*}
{takes into account the different number of hopping terms if $c_{ij}$ is purely real or imaginary.}

Because the Hamiltonian $H_q'$ satisfies statement \ref{thm:p2}, statement \ref{thm:p1} implies the existence of a fermion-qubit mapping $\m{m}': \mathcal{H}_f \rightarrow \mathcal{H} < \mathcal{H}_2^{\otimes (n+1)}$,
where $\mathcal{H}$ is the $(2^n)$-dimensional subspace of the $(n+1)$-qubit register stabilized by $Z_\mathcal{I} \otimes Z_n$, that gives the unitary equivalence $H_f = \m{m}' \cdot H_q' \cdot (\m{m}')^\dagger$. Thus, a quantum computer may use $H_q'$ to study $H_f$ instead of $H_q(\sigma)$, reducing the total and maximum Pauli weight cost of simulation at the expense of a single additional qubit.

Subsequent ancilla qubits can be added by renaming $H_q'$ to $H_q(\sigma)$ and iteratively applying the procedure described above. In introducing the $k$th ancilla qubit, choose a qubit subset ${\mathcal{I}_k}$ and define $\kappa_{ij}$ as in \eqref{eq:kappas}, with $Z_{\mathcal{I}_k}$ acting on the first $n$ qubits and the single-qubit Pauli acting on the $k$th ancilla qubit, leaving the previous ancilla qubits untouched. Perform the conditional modification of the hopping terms via Option 1 and Option 2 to produce an $(n{+}k)$-qubit Hamiltonian with lower Pauli weights.
Continue to add ancilla qubits incrementally until exhausting the available supply, or if there is no further reduction in the total or maximum Pauli weight of the hopping terms in the modified qubit Hamiltonian.

Naturally, this leads to the question of whether it is possible to find an optimal arrangement of multiple {qubit subsets $\mathcal{I}_0, \mathcal{I}_1, \dots \mathcal{I}_{p-1}$. While there was no requirement for the sets $\mathcal{I}_k$ to be pairwise disjoint, making this slight restriction simplifies the scenario in which both $Z_{\mathcal{I}_j}$ and $Z_{\mathcal{I}_k}$ are applied to the same hopping term in repeated instances of Option 1.} {This non-intersecting scenario} yields the optimization problem
\begin{align}
\begin{split}
\min_{\mathclap{\mathcal{I}_0, \ldots, \mathcal{I}_{p-1}}}\quad &\sum_{k=0}^{p-1} \sum_{(i,j) \in E_{\sigma}}\alpha_{ij}\min\Big(\lvert \{i,j\} \cap \mathcal{I}_k\rvert \bmod 2, d_{ij}(\mathcal{I}_k)\Big)\\
   \text{st}\quad & \forall i,j \in [p] : \mathcal{I}_i \cap \mathcal{I}_j = \emptyset \nonumber\, .
\end{split}
\end{align}
{This is problem that we solve in Section \ref{sec:ancillaresults}.}

\subsection{Constant-ancilla method for other mappings}
Our method for constructing constant-ancilla fermion-qubit mappings in Section \ref{sec:constantancillas} begins by using the Jordan--Wigner transformation to generate the base ancilla-free Hamiltonian in \eqref{eqn:jwancil}.
If we had instead used a different linear encoding, we predict that it would generally be computationally hard to devise an effective set of commuting Pauli strings $P_0,P_1,\dots, P_{p-1}$ for incremental ancilla addition because number, hopping and interaction terms map to Pauli strings with less obvious nonlocal structure than the $Z_{i+1} \dots Z_{j-1}$ factors of the Jordan--Wigner transformations. In modifying the Hamiltonian terms from $H_q(\sigma)$ to $H_q'$ with the addition of the $k$th ancilla qubit,
the protocol would need to search for Pauli operators $P_k = X_{\mathcal{I}_k} Y_{\mathcal{J}_k} Z_{\mathcal{K}_k}$ such that multiplying the terms of $H_q(\sigma)$ by $P_k$ reduces the total or maximum Pauli weight. As discussed elsewhere \cite{steudtner2019quantum, Chiew2023discoveringoptimal} a new requirement emerges that $[P_j,P_k]=0$ for all $j \in [k]$ in order to satisfy statement \ref{thm:p2}. 
Again, the reason this was straightforward for the Jordan--Wigner case was because we knew \textit{a priori} that the problematic parts of the hopping terms would consist only of $Z$ matrices. Nevertheless, we see this as a worthwhile line of inquiry because of the potential for other mappings to reduce the weight of qubit Hamiltonians from the outset.

\begin{table*}[btp]
\centering
\caption{A comparison of the Pauli weight for Jordan--Wigner, Bravyi--Kitaev, Parity Basis and Ternary Tree mappings of fermionic systems with Hamiltonian graphs that are square grids of size 3{$\times$}3 to 15{$\times$}15. Orderings are optimized for total Pauli weight, max Pauli weight, and total Pauli weight assuming real valued coefficients and the absence of operator terms, corresponding to the Fermi--Hubbard model.}
\label{tab:lattice}
\renewcommand{\arraystretch}{1.2}
\begin{tabular}{cc|ccccccccccccc}
\toprule
 & \(\lvert V\rvert\)   
     & 9       & 16       & 25       & 36       & 49       & 64       & 81       & 100      & 121      & 144      & 169      & 196       & 225 \\
\cmidrule(lr){1-15}
\multirow{4}{*}{\makecell{Total Pauli weight \\ $\mathcal{D} = \text{Num} + \text{ReHop} $ \\ $ + \text{ ImHop} + \text{Inter}$ }}
& BK & 304     & 635      & 1107     & 1712     & 2473     & 3331     & 4467     & 5741     & 7127     & 8850     & 10438    & 12595     & 14522 \\
& JW & \B{237} & \B{512}  & \B{909}  & \B{1460} & \B{2189} & \B{3104} & \B{4277} & 5632     & 7389     & 9320     & 11609    & 14364     & 17601 \\
& PB & 301     & 645      & 1147     & 1815     & 2683     & 3775     & 5155     & 6751     & 8729     & 10987    & 13657    & 17449     & 20505 \\
& TT & 313     & 628      & 1080     & 1676     & 2375     & 3237     & 4303     & \B{5473} & \B{6799} & \B{8342} & \B{9853} & \B{11844} & \B{13942}\\
\cmidrule(lr){1-15}
\multirow{4}{*}{\makecell{Maximum Pauli weight \\ $\mathcal{D} = \max(\text{Num},$ \\ $\text{ReHop}, \text{ImHop}, \text{Inter})$}}  
& BK & 5       & 7      & 9     & 9     & 11    & 11    & 12    & 13    & 13    & 13      & 14     & 15     & 15  \\
& JW & \B{4}   & \B{5}  & \B{6} & \B{7} & \B 8  & 9     & 11    & 12    & 14    & 15      & 16     & 18     & 20  \\
& PB & 5       & 6      & 7     & 8     & 9     & 10    & 12    & 13    & 14    & 15      & 17     & 18     & 20  \\
& TT & 5       & \B{5}  & 7     & \B{7} & \B 8  & \B 8  & \B 9  & \B 9  & \B 9  & \B{10}  & \B{10} & \B{10} & \B{11}\\
\cmidrule(lr){1-15}
\multirow{4}{*}{\makecell{Fermi--Hubbard \\ $\mathcal{D} = \text{ReHop} + \text{Inter}$}}   
& BK & 178      & 372     & 647     & 1005      & 1458      & 1993      & 2655     & 3445     & 4277     & 5205     & 6207     & 7578     & 8811  \\
& JW & \B{138}  & \B{296} & \B{522} & \B{832}   & \B{1238}  & \B{1746}  & \B{2392} & \B{3142} & 4088     & 5190     & 6388     & 7850     & 9936  \\
& PB & 180      & 386     & 683     & 1075      & 1585      & 2239      & 3033     & 3966     & 5126     & 6369     & 8079     & 9891     & 12937 \\
& TT & 186      & 368     & 638     & 990       & 1412      & 1918      & 2550     & 3237     & \B{4038} & \B{4938} & \B{5887} & \B{6869} & \B{8086} \\
\bottomrule
\end{tabular}
\end{table*}

\begin{table}[btp]
\centering
\caption{Total Pauli weight $(\mathcal{D} = \text{Num} + \text{ReHop} + \text{ImHop} + \text{Inter}$) for fermionic systems of various Hamiltonian graphs with 64 vertices using the optimized fermionic label ordering for the Jordan--Wigner and Ternary Tree transformations.}
\label{tab:graphs}
\begin{tabular}{rcc}
\toprule
64-mode system graph   & JW & TT\\
\cmidrule(lr){1-3}
Hex-Lattice           & 7564  & 5489   \\
Tri-Lattice           & 2384  & 2478   \\
Periodic Hex-Lattice  & 8584  & 4794   \\
Periodic Tri-Lattice  & 2704  & 2356   \\
Random 3-Regular      & 2888  & 2245   \\
Margulis Gabber Galil & 11784 & 6543   \\
Chordal Cycle         & 6976  & 4431 \\
\bottomrule
\end{tabular}
\end{table}

\section{Experiments} \label{sec:experiments}

In this section, we compare the quality of results for the Jordan--Wigner, Bravyi--Kitaev, Parity Basis, and Ternary Tree encodings in multiple contexts, including on arbitrary lattice geometry and lattices of very large size. What's more, we show that although the total {and maximum} Pauli weight of hopping terms for the Jordan--Wigner encoding may fall behind other options, the use of {a small number of ancilla qubits} 
is able to produce results comparable or better to these other options.

All optimizations are performed using the Hexaly library on an AMD Ryzen 5 7640U. Unless specified, all coefficients are assumed to have non-zero real and imaginary parts.

\begin{figure}
    \centering
    \begin{minipage}{\linewidth}
        \begin{minipage}{0.1\linewidth}
            \subcaption{}
            \label{fig:stabilizer_bk}
        \end{minipage}
        \begin{minipage}{0.8\linewidth}
            \centering
            \includegraphics[width=\linewidth]{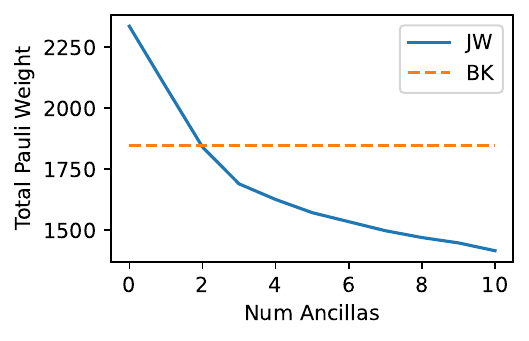}
        \end{minipage}
    \end{minipage}
    \vspace{1em}
    \begin{minipage}{\linewidth}
        \begin{minipage}{0.1\linewidth}
            \subcaption{}
            \label{fig:stabilizer_tt}
        \end{minipage}
        \begin{minipage}{0.8\linewidth}
            \centering
            \includegraphics[width=\linewidth]{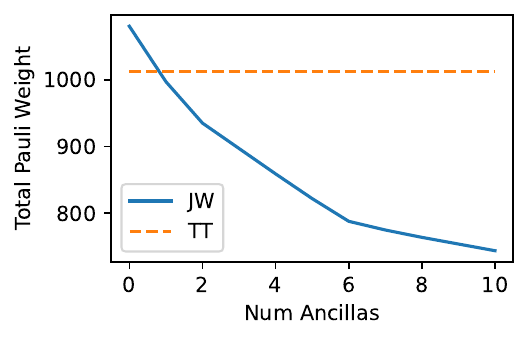}
        \end{minipage}
    \end{minipage}
    \caption{
    Comparison of the total Pauli weight of hopping terms ($\mathcal{D} = \text{ReHop} +\text{ImHop}$) corresponding to Jordan--Wigner with an increasing number of {ancilla qubits} against Bravyi--Kitaev (\subref{fig:stabilizer_bk}) and Ternary Tree (\subref{fig:stabilizer_tt}) for the 8{$\times$}8 and 5{$\times$}8 grid respectively. }
    \label{fig:stabilizer}
\end{figure}

\subsection{{Encoding comparison with optimized fermionic order}} \label{sec:1sumres}

In this section, we examine the difference in {the total and maximum} Pauli weights yielded by the four primary encodings.
In Table \ref{tab:lattice}, we compare the results of the Jordan--Wigner, Bravyi--Kitaev, Parity Basis and Ternary Tree transformations applied to a series of square grids of increasing size. We show results for three different optimization objectives: total Pauli weight, maximum Pauli weight, and total Pauli weight for a model with only real-valued coefficients and no operator terms, which corresponds to the popular Fermi--Hubbard model \cite{PhysRevLett.10.159}. Here, we see that Jordan--Wigner is the preferred encoding for smaller grids, allowing orderings which produce a small maximum and total Pauli weight compared to other methods.  However, for all but the smallest input, Ternary Tree remains comparable, and eventually emerges as the superior ordering. Bravy--Kitaev's better scaling similarly allows it to outperform Jordan--Wigner for larger grids. Within the Jordan--Wigner orderings, we recover the optimal Mitchison--Durbin ordering \cite{mitchison1986optimal} for square lattices of size 6$\times$6 and larger, agreeing with the analytical result in \cite{Chiew2023discoveringoptimal}.

This result is supported by our findings in Table \ref{tab:graphs}, which shows the best observed encoding and total Pauli weight for a number of different graph structures. Ternary Tree performed best for the vast majority of tested instances, with the only exception being Jordan--Wigner for the comparably simple triangular lattice.

\subsection{{Jordan--Wigner transformations with optimal ancilla use}} \label{sec:ancillaresults}

In this section, we examine the improvement to {the Jordan--Wigner encoding scheme} provided by {our method of incrementally adding ancilla qubits}. All experiments provided in this section concern themselves only with the total Pauli weight of the hopping terms, i.e.\ $\mathcal{D} = \text{ReHop} + \text{ImHop}$. 

In Figure \ref{fig:stabilizer}, we compare Jordan--Wigner with multiple {ancilla qubits} with the {ancilla-free} Bravyi--Kitaev and Ternary Tree transformations on the 8{$\times$}8 and 5{$\times$}8 grids, respectively. These graphs are chosen so that the traditional definitions from \eqref{eq:bk} and \eqref{eqn:tt} hold, respectively. Here, we see that adding only a few {ancilla qubits} can significantly improve the Pauli weights associated with the Jordan--Wigner transformation, to the point where it overtakes the Pauli weight of Ternary Tree. In the case of the 8{$\times$}8 grid, using 10 {ancilla qubits} allows us to reduce the Pauli weight of the {hopping terms} in the final {Jordan--Wigner} Hamiltonian by 39.3\% {compared to the ancilla-free Jordan--Wigner Hamiltonian}.

In Table \ref{tab:stab}, we compare the Jordan--Wigner transformation with zero and ten ancilla qubits to the {ancilla-free} Ternary Tree transformation on a number of different graph structures. With ten ancilla qubits, the Pauli weight cost of hopping terms for the Jordan--Wigner encoding can be reduced by as much as 66.9\%. The resulting hopping terms have a total Pauli weight comparable  to that of the Ternary Tree transformation, or less.

\begin{table}[btp]
\centering
\caption{Total Pauli weights of hopping terms $(\mathcal{D} = \text{ReHop} + \text{ImHop})$ for fermionic systems of various Hamiltonian graphs with 64 vertices using Jordan--Wigner, Ternary Tree, and Jordan--Wigner with 10 ancilla qubits.}
\label{tab:stab}
\begin{tabular}{rccc}
\toprule
 64-mode system graph                     & JW    & TT   & JW + 10 ancillas \\
\cmidrule(lr){1-4}
Hex-Lattice           & 6672  & 3871 & 3757          \\
Tri-Lattice           & 1952  & 1748 & 1253          \\
Periodic Hex-Lattice  & 7800  & 3429 & 3487          \\
Periodic Tri-Lattice  & 2320  & 1664 & 1150          \\
Random 3-Regular      & 2504  & 1599 & 1253          \\
Margulis Gabber Galil & 10792 & 4698 & 3567          \\
Chordal Cycle         & 6210  & 3213 & 2445         \\
\bottomrule
\end{tabular}
\end{table}

\section{Conclusion}

We have introduced a cost model for
the total and maximum Pauli weight of the output of linear encodings, a type of ancilla-free fermion-qubit mapping, and have found optimized fermionic orderings for several popular transformations using quadratic assignment. Empirically, these optimized orderings are able to recover known optimal solutions in the case of the Jordan--Wigner mapping. Moreover, we find an optimal strategy to cancel out disjoint strings of Pauli-$Z$ gates, utilizing 
ancilla qubits to reduce the Pauli weight of qubit Hamiltonians associated with the Jordan--Wigner transformation. According to our results, the latter strategy requires only of 10 ancilla qubits to reduce Hamiltonian Pauli weight by as much as 67\%, outperforming the optimized ancilla-free Ternary Tree transformation. Future work might expand our approach to more complex cost functions that incorporate gate depth and circuit compilation, and generalise our incremental ancilla method to mappings beyond the Jordan--Wigner transformation.

\bibliographystyle{IEEEtran}
\bibliography{refs}

\end{document}